\documentclass[aps,prl,twocolumn,groupedaddress]{revtex4}
\usepackage{amsmath}
\usepackage{graphics}
\usepackage{graphicx}
\usepackage{epsfig}
\usepackage{amssymb}
\usepackage{amsmath}
\usepackage{amsfonts}

\begin{document}

\title{Asymmetric superconductivity in metallic systems}
\author{Mucio A. \surname{Continentino} and Igor T. \surname{ Padilha }}
\affiliation{Instituto de F\'{\i}sica, Universidade Federal
Fluminense, Campus da Praia Vermelha,  Niter\'oi, RJ, 24.210-340,
Brazil} \email{mucio@if.uff.br}
\date{\today}

\begin{abstract}
Different types of superfluid ground states have been investigated
in systems of two species of fermions with Fermi surfaces that do
not match. This study is relevant for cold atomic systems, condensed
matter physics and quark matter. In this paper we consider this
problem in the case the fermionic quasi-particles can transmute into
one another and only their total number is conserved. We use a BCS
approximation to study superconductivity in two-band metallic
systems with inter and intra-band interactions. Tuning the
hybridization between the bands varies the mismatch of the Fermi
surfaces and produces different instabilities. For inter-band
attractive interactions we find a first order normal-superconductor
and a homogeneous metastable phase with gapless excitations. In the
case of intra-band interactions, the transition from the
superconductor to the normal state as hybridization increases is
continuous and associated with a quantum critical point. The case
when both interactions are present is also considered.
\end{abstract}

\maketitle


\section{Introduction}

A new superfluid ground state originally named interior gap or
breached pairing (BP) superfluidity has been recently investigated
\cite{liu,liu2,liu3}. This state presents a homogeneous mixture of
normal and superfluid properties and should occur in fermionic
systems with different Fermi surfaces. Superfluidity develops at
the Fermi surface of the quasi-particles with the smallest Fermi
momentum. Since its proposal much work has been done in
understanding the nature of this state and in particular its
stability \cite{caldas,liu2}. In this work we consider the
possibility of interior gap superfluidity in systems where the
quasi-particles can transmute into one another and only their
total number is conserved. Our results are directly relevant for
condensed matter systems, cold-atom systems \cite{nature} in the
presence of Rabi coupling \cite{liu3} and should be of interest
for the study of color superconductivity on the core of neutron
stars with quarks that can interchange their flavors
\cite{caldas,livro,alford1}. For concreteness we focus in the
former problem. Specifically, on superconductivity in transition
metals (TM) or rare earth inter-metallic systems where a large
$a$-band
of conduction electrons ($s$, or $p$) coexist with a narrow $b$%
-band of $d$ or $f$-electrons. We consider inter and intra-band
attractive interactions. In both cases we show that a finite
interaction is necessary to give rise to superconductivity,
differently from the Bardeen-Cooper-Schrieffer (BCS) \cite{bcs}
case. For inter-band attraction the transition into the
superconducting state is first order. We find a new superconducting
state with features of the internal gap or breached pairing state
\cite{liu2} including {\em Fermi surfaces} with gapless excitations.
In the intra-band case there is a superconducting quantum critical
point (QCP) that can be probed in experiments under pressure.
Finally, we include both inter and intra-band interactions and show
that in this case gapless excitations are generally suppressed.

\section{Inter-band superconductivity}

We consider initially a model with two types of quasi-particles, $a$
and $b$, with an attractive interaction \cite{kondo} $g$ and a
hybridization term $V$ that mixes different quasi-particles states.
This one-body mixing term $V$ arises from overlap of different
orbitals either in the same, or different sites. It is a useful
control parameter since it can be varied by external pressure
allowing to explore the phase diagram and quantum phase transitions
of the model. The Hamiltonian is given by,
\begin{align}
& H=\sum_{k\sigma }\epsilon _{k}^{a}a_{k\sigma }^{\dag }a_{k\sigma
}+\sum_{k\sigma }\epsilon _{k}^{b}b_{k\sigma }^{\dag }b_{k\sigma }+
\label{lw} \\
& g\sum_{kk^{\prime }\sigma }a_{k^{\prime }\sigma }^{\dag }b_{-k^{\prime
}-\sigma }^{\dag }b_{-k-\sigma }a_{k\sigma }+\sum_{k\sigma }V_{k}(a_{k\sigma
}^{\dag }b_{k\sigma }+b_{k\sigma }^{\dag }a_{k\sigma })  \notag
\end{align}%
where $a_{k\sigma }^{\dag }$ and $b_{-k^{\prime }-\sigma }^{\dag
}$ are creation operators for the light $a$ and the heavy
$b$-quasi-particles, respectively. The index $\ell =a,b$. The
dispersion relations $\epsilon _{k}^{\ell }=k^{2}/2m_{\ell }-\mu
_{\ell }$ and the ratio between effective masses is taken as
$\alpha = m_{a}/m_{b} <1$. When $V=0$ this model requires a
critical value $\Delta _{ab}^{c}$ of the order parameter, $\Delta
_{ab}=-g\sum_{k}<a_{k}b_{-k}>$, to sustain BCS superconductivity
\cite{liu} (we neglect spin indexes here). The instability of the
BCS phase for $\Delta _{ab}<\Delta _{ab}^{c}$ is
associated with a soft mode at a wave-vector $k_{c}$ ($%
k_{F}^{a}<k_{c}<k_{F}^{b}$) which suggests a transition to a Fulde
and Ferrel, Larkin, Ovchinnikov (FFLO) state \cite{fflo} with a
characteristic wave-vector $k=k_{c}$. However the window of
parameters for which this phase is stable is very narrow
\cite{izuyama} and a BP or Sarma phase \cite{liu,sarma} has also
been considered. Since this corresponds to a maximum of the free
energy, a mixed phase with normal and superconducting regions
\cite{caldas} was proposed as an alternative ground state for
$\Delta _{ab}<\Delta _{ab}^{c}$.

\begin{figure}[tbh]
\centering{\ \includegraphics[scale=0.7]{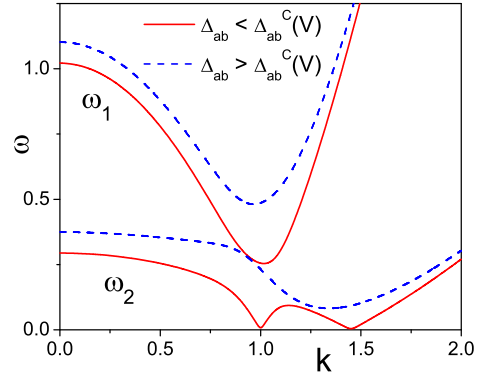}}
\caption{(Color online) Dispersion relations for $V=0.1$: $\Delta
_{ab}=0.1<\Delta _{ab}^{c}(V=0.1)\sim 0.224$ (full line) and
$\Delta _{ab}=0.35>\Delta _{ab}^{c}(V=0.1)$ (dashed line). }
\label{fig1}
\end{figure}

In order to obtain the spectrum of excitations of Eq.\ref{lw}
within the BCS (mean-field) approximation, we use the equation of
motion method to calculate standard and anomalous Greens
functions. Excitonic type of correlations that just renormalize
the hybridization \cite{sarasua} have been neglected. The order
parameter $\Delta _{ab}$ is obtained self-consistently from the
anomalous Greens function,
\begin{equation}
\ll a_{k};b_{-k}\gg =\frac{-\Delta _{ab}\left[ (\omega -\epsilon
_{k}^{b})(\omega +\epsilon _{k}^{a})+(V^{2}-\Delta _{ab}^{2})\right] }{%
(\omega ^{2}-\omega _{1}^{2})(\omega ^{2}-\omega _{2}^{2})}.  \label{eqn1}
\end{equation}
Besides, hybridization combined with the interaction $g$ can give
rise to a net attraction between the $b$ quasi-particles, even in
the absence of such interaction in the original Hamiltonian. This
becomes manifest in the calculations where we find a finite
anomalous Greens function $\ll b_{k};b_{-k}\gg $ given by,
\begin{equation}
\label{bb} \ll b_k;b_{-k}\gg=\frac{-2 \Delta_{ab} V
\epsilon_k^a}{(\omega^2-\omega_1^2)(\omega^2- \omega_2^2)}
\end{equation}
It turns out however that the anomalous correlation function
$<b_{k}b_{-k}>$ is identically zero in the present calculation.
The poles of the Greens function occur for $\omega =\pm \omega
_{12}(k)$, where,
\begin{equation}
\omega _{12}(k)=\sqrt{A_{k}\pm \sqrt{B_{k}}}  \label{poles}
\end{equation}%
with,
\begin{equation}
\label{ak}
A_{k}=\frac{(\epsilon
_{k}^{a2}+\epsilon_{k}^{b2})}{2}+(V^{2}+\Delta _{ab}^{2})
\end{equation}%
and
\begin{equation}
\label{bk}
B_{k}=\frac{(\epsilon _{k}^{a2}-\epsilon
_{k}^{b2})^{2}}{4}+(\epsilon _{k}^{a}+\epsilon
_{k}^{b})^{2}V^{2}+4V^{2}\Delta _{ab}^{2}+(\epsilon
_{k}^{a}-\epsilon _{k}^{b})^{2}\Delta _{ab}^{2}
\end{equation}
In the
calculations below we take $\hbar^2/(2m_a \mu_a)=1$ since the
relevant parameter is the mass ratio $\alpha$. Energies are
normalized by the Fermi energy $\mu _{a}$ of the light
quasi-particles, such that, in all figures the quantities in the
axis are numbers. The original band dispersion relations are then
written as, $\epsilon _{k}^{a}=k^{2}-1$ and $\epsilon
_{k}^{b}=\alpha k^{2}-b$. Assuming all states with negative energy
are
filled, we have $k_{F}^{a}=1$. We take $k_{F}^{b}=1.45$, $%
\alpha =1/7$, such that, $\mu _{b}/\mu _{a}=b\approx 0.30$ as in Ref. \cite%
{caldas} for cold atomic systems \cite{nota}. These numbers are also
appropriate to describe transition metals (TM) for which typical
values of the bandwidths ($\mu_{a,b}$) are a few electronvolts with
$g$ and $V$ both of order $10^{-1}$ or $10^{-2}$. The mass ratio
$\alpha$ ranges from $10^{-1}$ for TM to $10^{-3}$ for heavy
fermions (HF) \cite{livroM}. The general features of the solutions
we obtain are however independent of a particular set of parameters.
Figure \ref{fig1} shows the dispersion relations of the excitations.
Differently from the case $V=0$, there are no negative values of the
energy \cite{liu} for any $\Delta _{ab}\neq 0$. However, the
dispersion relations vanish at two, two-dimensional {\em Fermi
surfaces}\cite{alford} determined by,
\begin{equation}
\epsilon_k^a \epsilon_k^b +(\Delta_{ab}^2-V^2)=0 \label{fs}
\end{equation}%
for $\Delta _{ab}\leq \Delta _{ab}^{c}(V)$ where,
\begin{equation}
\label{critical} \Delta_{ab}^{c}(V)=\sqrt{\Delta
_{ab}^{c}(V=0)^{2}+V^{2}}
\end{equation}
with \cite{caldas} $\Delta _{ab}^{c}(V=0)=|(\alpha
-b)|/2\sqrt{\alpha }$. As $\Delta _{ab}$, i.e., the coupling g
increases and reaches $\Delta _{ab}^{c}(V)$, the two gapless Fermi
surfaces (FS) merge at a critical FS.  For $\Delta _{ab} > \Delta
_{ab}^{c}(V)$ the dispersion relations are BCS-like with a finite
gap for excitations (see Fig. \ref{fig1}). The instability of the
BCS phase can also be triggered by the hybridization which
increases the mismatch of the Fermi surfaces due to a {\it
repulsion} between the bands \cite{livro}. It occurs at a critical
value, $V_c= \sqrt{\Delta_{ab}^2-\alpha({k_F^b}^2-{k_F^a}^2)/4}$
for a fixed $\Delta_{ab} > \sqrt{\alpha({k_F^b}^2-{k_F^a}^2)/4}$.
Both instabilities, due to increasing hybridization, or by
decreasing the coupling $g$ (or $\Delta_{ab}$), belong to the same
universality class and are associated with a soft mode at a
wave-vector $k_c$.

Dispersion relations with similar features of those shown in
Fig.\ref{fig1} were obtained for color superconductivity
\cite{alford}. An additional p-wave instability at the new
FS\cite{new}, which is outside the scope of the present mean-field
approach, has been investigated. In the metallic problem there is
the possibility of additional pairing in the s-wave channel of the
same type of particles due to the extra spins degree of freedom
(see Eq. \ref{bb}). However, as pointed out before, the relevant
anomalous correlation function associated with this Greens
function turns out to be identically zero. Notice that the
dispersion of the fermions close to the new FS are linear and at
least in d=2, this requires a finite interaction for pairing to
occur \cite{marino}. It would be interesting to consider other
types of instability at these Fermi surfaces, as spin density wave
ordering.

\begin{figure}[th]
\centering{\ \includegraphics[scale=0.7]{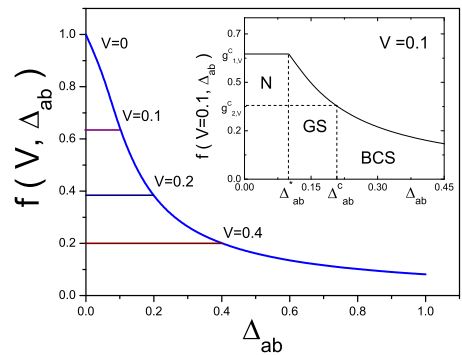}} \caption{(Color
online) Gap function $f$ normalized to its value at $V=0$, for
different values of hybridization $V$. The inset shows the phases
associated with different values of the order parameter $\Delta
_{ab}$ for a fixed hybridization $V=0.1$. N is a normal phase, GS
and BCS correspond to gapless and BCS
superconducting phases, respectively. The interactions $g_{1,V}^{c}$ and $%
g_{2,V}^{c}$ mark the limits of the gapless (GS) and BCS
superconducting phases (see Fig.\protect\ref{fig3}).} \label{fig2}
\end{figure}

From the discontinuity of the Greens functions on the real axis we
can obtain the anomalous correlation function characterizing the
superconducting state. The self-consistent equation for the order
parameter $\Delta _{ab}=-g\sum_{k}<b_{-k}a_{k}>$ at $T \ne 0$ is
given by,
\begin{equation}
\frac{1}{g}\!=\!\sum_{j=1}^{2}\!\!\int \!\!\frac{d^{3}k}{(2\pi )^{3}}\!\!%
\left[ \frac{(-1)^{j}}{2\sqrt{B_{k}}}\!\left( \frac{\omega
_{j}(k)^{2}\!-\!E^{2}(k)}{2\omega _{j}(k)}\!\right) \!\tanh (\frac{\beta
\omega _{j}(k)}{2})\!\right]   \label{op}
\end{equation}%
where $E^{2}(k)=\epsilon _{k}^{a}\epsilon _{k}^{b}+(\Delta
_{ab}^{2}-V^{2})$. This equation can be written as, $1/g\rho
=f(V,\Delta _{ab})$, where $\rho$ is the density of states at the
Fermi level of the $a$-band. The function $f(V,\Delta _{ab})$ is
plotted in Fig.\ref{fig2} for several values of the
hybridization parameter. For $V=0$ a solution with a finite order parameter $%
\Delta _{ab}$ only exists for $(1/g)<(1/g_{1}^{c})=\rho f(0,0)$ with $%
f(0,0)=(2/(1-\alpha ))|\ln [(b-\alpha )/(\omega _{c}(1-\alpha )+(b-\alpha
))]|\sim 0.123$. The quantity $\omega _{c}=0.01$ is a small cut-off energy
around the Fermi energy where the integrals in energy are performed. Still
for $V=0$ there is another characteristic value of the coupling $%
(1/g_{2}^{c})=\rho f(0,\Delta _{ab}^{c}(V=0))$, such that, for
$g_{1}^{c}<g<g_{2}^{c}$ the system presents a BP or a mixed phase
\cite{caldas}. For $g>g_{2}^{c}$ superconductivity is of the BCS
type \cite{caldas}. Since the BP phase appears as a maximum of the
free energy, an alternative state for $g_{1}^{c}<g<g_{2}^{c}$ is a
mixed phase with coexisting normal and superconducting BCS-like
regions \cite{caldas}. For $g>g_{2}^{c}$ the superconducting BCS
is the stable ground state \cite{caldas}.

\begin{figure}[h]
\centering{\ \includegraphics[scale=0.7]{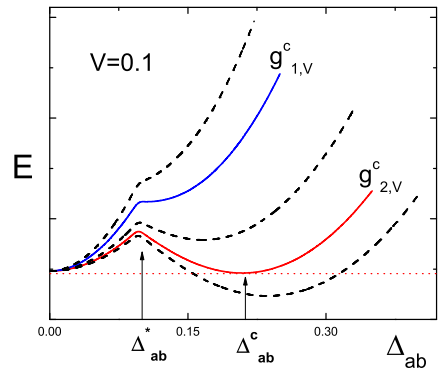}}
\caption{(Color online) Free energy at zero temperature as a
function of the order parameter for different values of the
interaction $g$ and a fixed hybridization $V=0.1$. For
$g_{2,V}^{c}>g>g_{1,V}^{c}$ there is a metastable superconducting
(GS) phase with $\Delta _{ab}^{c}(V)>\Delta _{ab}>\Delta
_{ab}^{\ast }(V)$ and gapless excitations. } \label{fig3}
\end{figure}

As hybridization is turned on at zero temperature a stronger value
of the coupling $g$ is necessary to obtain a superconducting
solution, since $f(V,0)<f(0,0)$ (Fig.\ref{fig2}). The function
$f(V,\Delta_{ab} )$ normalized by its value for $V=0$ is shown in
Fig.\ref{fig2}. Although hybridization acts in detriment of
superconductivity we notice that, at least for small values of
$V$, a weak coupling approximation is still justified, as for
$V=0$ treated in Ref. \cite{liu}. The function $f(V,\Delta_{ab})$
is flat up to $\Delta _{ab}=\Delta _{ab}^{\ast }(V)\sim V$ (see
Fig.\ref{fig2}), such that, when the coupling $g$ is strong enough
to stabilize a superconducting solution it occurs already at a
finite value of the order parameter. Consequently, for $V\neq 0$
the quantum, normal to superconducting
phase transition as a function of the coupling $g$ is first order. For $%
\Delta _{ab}^{\ast }(V)<\Delta _{ab}<\Delta _{ab}^{c}(V)$ there is
a superconducting solution, the GS phase in Fig.\ref{fig2}, with
the
spectra of excitations shown in Fig.\ref%
{fig1} as full lines. This solution corresponds to a metastable
minimum of the free energy.
This is shown in Fig.\ref{fig3} where we plot the zero temperature
free energy for a fixed hybridization, $V=0.1$, and different
values of the coupling parameter $g$. The metastable minima appear
for $g_{2,V}^{c}>g>g_{1,V}^{c}$ and  occur at values of the order
parameter $\Delta _{ab}^{\ast }(V)<\Delta _{ab}<\Delta
_{ab}^{c}(V)$, as shown in Fig.\ref{fig3}. For these values of
$\Delta _{ab}$ the gaps in the lower branch of the dispersion
relations vanish at two two-dimensional {\em Fermi surfaces} (see
Fig.\ref{fig1}). This superconducting phase has similarities to
the BP superconductor \cite{liu} in that both have gapless
excitations, but with the difference that the present one
corresponds to a minimum, even though metastable, of the free
energy.
At $g=g_{2,V}^{c}$ the normal and superconducting phase exchange
stability at a quantum first order phase transition. The critical
value $g_{2,V}^{c}$ for a fixed $V$  is given by the condition
$E[\Delta_{ab}(V)=0, g=g_{2,V}^{c}]=E[\Delta_{ab}(V),
g=g_{2,V}^{c}]$ where $E(\Delta_{ab}, V, g)$ is the zero
temperature free energy. As the coupling $g$ increases beyond
$g_{1,V}^{c}$, the first solution for this equation is obtained
for $\Delta_{ab}(V)=\Delta_{ab}^{c}(V)$ (see Eq. \ref{critical}
and Fig. \ref{fig3}). Thus, the first order transition {\it as a
function of the coupling strength} occurs together with the change
in the excitation spectrum. For $g>g_{2,V}^{c}$ the stable ground
state is a BCS type of superconductor with gapped excitations
since the stable free energy minimum occurs for values of the
order parameter $\Delta _{ab}>\Delta _{ab}^{c}(V)$ (see
Fig.\ref{fig3}). The dispersion relations are like those shown as
dashed lines in Fig.\ref{fig1}. We point out that for $g \le
g_{1,V}^{c}$ ($\Delta \le \Delta _{ab}^{\ast }(V)$) the metastable
minimum of the free energy disappears (Fig.\ref{fig3}). Then, the
value $g=g_{1,V}^{c}$ marks the limit of stability of the BCS-like
superconducting phase into the normal phase.  The other limit, of
the metastable normal phase in the superconducting phase is not
shown. Then, as one increases hybridization in a two-band BCS
superconductor with attractive inter-band interactions, two main
effects take place. First, hybridization increases the mismatch
between the Fermi surfaces giving rise to a first order transition
from the BCS-superconductor to the normal state. At this
transition appears a metastable GS phase with two two-dimensional
Fermi surfaces with gapless excitations. Differently from the
breached pairing state, in this GS phase pairing takes place among
quasi-particles with momenta between $k_F^a$ and $k_F^b$. The
mixing of the quasi-particles allow them to take advantage of the
condensation energy in this range of k-space reducing the energy
of the GS phase with respect to the BP state.

\section{Intra-band interactions}

\begin{figure}[th]
\centering{\ \includegraphics[scale=0.7]{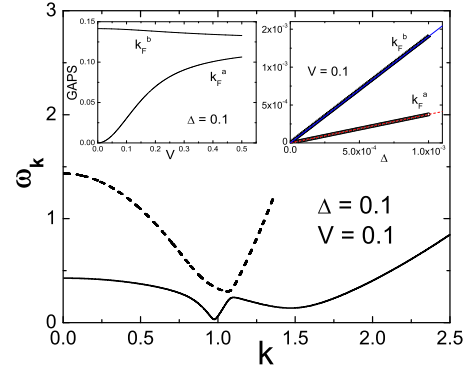}}
\caption{(Color online) Dispersion relations for model Eq.
\protect\ref{gloria}. Inset
shows the energy of the minima in the lower dispersion close to $k_F^a$ and $%
k_F^b$ as a function of $\Delta$ and $V$.} \label{fig4}
\end{figure}

Next we consider a closely related model which is relevant for
many physical systems of interest as inter-metallic compounds,
high $T_c$ and heavy fermion materials \cite{gloria}. It consists
of a narrow band of quasi-particles with an attractive interaction
that hybridizes with another band. The Hamiltonian is given by,
\begin{align}  \label{gloria}
& H = \sum_{k \sigma} \epsilon_{k}^{a}a^{\dag}_{k \sigma}a_{k \sigma}+
\sum_{k \sigma} \epsilon_{k}^{b}b^{\dag}_{k \sigma} b_{k \sigma} + \\
&g_b \sum_{k k^{\prime} \sigma}b^{\dag}_{k^{\prime} \sigma}
b^{\dag}_{-k^{\prime} -\sigma} b_{-k -\sigma} b_{k \sigma}+
\sum_{k \sigma} V_{k} (a^{\dag}_{k \sigma}b_{k \sigma}+b_{k
\sigma}^{\dag}a_{k \sigma}). \notag
\end{align}
In this case we have to keep track of the spin indexes since the
operators associated with the particles forming the pairs do not
necessarily anticommute. The dispersion relations of the
quasi-particles in the BCS approximation are obtained, as before,
from the poles of the Greens functions. They are given by,
$\omega_{12}(k)=\sqrt{\tilde{A}_k \pm\sqrt{\tilde{B}_k}}$ with,
\begin{equation}
\tilde{A}_k=\frac{\epsilon_k^{a 2}+\epsilon_k^{b 2}}{2}+V^2+
\frac{\Delta^2}{2}
\end{equation}
and
\begin{equation}
\tilde{B}_k = (\frac{\epsilon_k^{b 2}-\epsilon_k^{a 2}+\Delta^2}{2})^2 +  V^2 \left[%
(\epsilon_k^{a}+\epsilon_k^{b})^2 + \Delta^2 \right]
\end{equation}
where $\Delta=-g_b \sum_k <b_{-k \uparrow}b_{k \downarrow}>$ is a
new order parameter associated with superconductivity in the
narrow b-band. For $V \ne 0$, the dispersion relations above do
not vanish for any value of $k$, as can be verified from the
condition,
\begin{equation}
Z(k)=\tilde{A}^2_k-\tilde{B}_k=(\epsilon_k^{a}
\epsilon_k^{b}-V^2)^2 +\Delta^2\epsilon_k^{a 2}=0
\end{equation}
which has no real solution. These new dispersions are shown in Fig.%
\ref{fig4}. The lower branch of the dispersion has dips for
wave-vectors close to the original Fermi wave-vectors. The gaps at
the dips vary linearly with the order parameter $\Delta$, for
fixed $V$, as shown in the inset. This suggests that the modes at
the dips behave as roton-like excitations with a roton gap
proportional to the superconducting order parameter. For fixed
$\Delta$ changing the hybridization, the gap close to $k_F^a$ can
become arbitrarily small (inset of Fig. \ref{fig4}). As shown in
this figure this gap may be smaller than the gap at $k_F^b$
associated with superconductivity. This has experimental
consequences as the activated behavior of thermodynamic properties
will be dominated by the smaller gap due to hybridization.
\begin{figure}[th]
\centering{\includegraphics[scale=0.7]{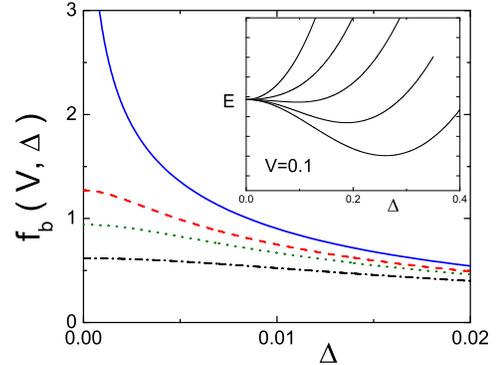}} \caption{(Color
online) Gap function $f_b(V, \Delta)$ for different values of
hybridization ($V=0.10$, $0.12$, $0.13$ and $0.15$ from top to
bottom). Inset: Free energy ($T=0$) as a function of the order
parameter for different values of the coupling $g_b$. As this
increases, the minimum moves continuously from $\Delta=0$ to a
finite value as the system enters in the superconducting phase.
Similar curves are obtained, but with the minimum moving to
$\Delta=0$, if $V$ is increased starting from $V_0$ for a fixed
$g_b>g^c_b(V_0)$.} \label{fig5}
\end{figure}
The gap equation at $T=0$ is given by,
\begin{equation}  \label{gapgloria}
\frac{1}{g_b \rho_b}\!\!=\!\!f_b(\Delta,\!
V)\!\!=\!\!\frac{1}{2}\!\!
\int_{-\omega_0}^{\omega_0}\!\!\!\!\! d\epsilon\! \frac{ 1}{%
\omega_1(\epsilon)\!+\! \omega_2(\epsilon)}\!\left[\!1\!\!+\!\! \frac{%
(\epsilon \!\!+\!(b\!-\!\alpha))^2}{\alpha^2 \sqrt{Z(\epsilon)}}\!
\right]
\end{equation}
where $\rho_b$ is the density of states of the narrow b-band at the
Fermi level. For $V=0$ this reduces to the BCS gap equation for a
single $b$-band. In Fig.\ref{fig5} we show $f_b(V, \Delta)$ as a
function of $\Delta$ for several values of the hybridization. We
find that $f_b(V, 0)$ is finite for values of $V \ne 0$ showing that
in this case a finite interaction $g^c_b(V)=1/(\rho_b f_b(V, 0))$ is
necessary for the appearance of superconductivity differently from a
single BCS-band.  Notice that for physical values of the
hybridization, $V \le 0.12$ the condition for superconductivity
$g^c_b(V)\rho_b < 1$ is still in the weak coupling regime (see Fig.
\ref{fig5}). Then for small but reasonable values of $V$ the present
BCS approach yields useful results. As in the previous section, we
get in this intra-band case a finite Greens function $\ll
a_{k\uparrow};b_{-k\downarrow}\gg$, but we find that the anomalous
correlation function $<b_{-k\downarrow}a_{k\uparrow}>$ is
identically zero.

The quantum phase transition at $g^c_b(V)$ is second order, as can
be seen from Fig.\ref{fig5}, since the condition $1/g^c_b(V)\rho_b
=f_b(V, \Delta))$ is first satisfied for $\Delta=0$. Besides the
free energy curves in the inset of this figure show directly the
continuous nature of the transition. Quantum fluctuations as
coupling to the electromagnetic field \cite{qforder} could
eventually drive this transition first order, but this is outside
the scope of the present BCS approximation. Since in real
multi-band systems some hybridization always occurs the existence
of a quantum critical point should be ubiquitous in
superconducting compounds with intra-band attractive interactions.
This QCP can be reached applying pressure in the system to vary
the overlap of the atomic orbitals and consequently $V$, as is
common, for example, in the study of HF materials \cite{livroM}.

\section{Intra and inter-band case}

\begin{figure}[th]
\centering{\includegraphics[scale=0.7]{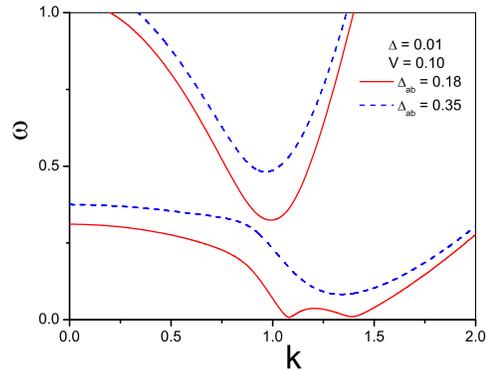}} \caption{(Color
online) Dispersion relations for the general case (intra and
inter-band attraction). We consider two cases of $\Delta_{ab}$
larger and smaller than $\Delta_{ab}^c(g_b=0)\approx 0.2$. In the
latter case, the dispersion relation can become very small for
wave-vectors close to the original Fermi surfaces.} \label{fig6}
\end{figure}

\begin{figure}[th]
\centering{\includegraphics[scale=1.3]{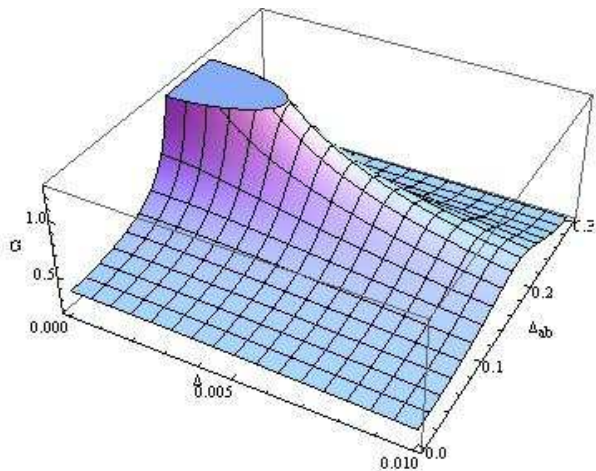}} \caption{(Color
online) The gap function \protect$\bar{G}(\Delta, \Delta_{ab})$ for
$V=0.15$. For small $\Delta$ there is a region of first order
transitions for $\Delta_{ab} \sim V$.} \label{fig7}
\end{figure}

Finally, we address the general case of attraction among the heavy
$b$-quasi-particles and the $a$ and $b$ fermions (inter and
intra-band attractive interactions). The calculations are long but
can be carried out analytically. The new excitations are obtained
from the equation,
\begin{align}
& \omega^4 - \left[\epsilon_k^{a2} + \epsilon_k^{b2} + 2( V^2 +
\Delta_{ab}^2) + \Delta^2 \right] \omega^2 + 4  V \Delta
\Delta_{ab}  \omega + \nonumber  \\
& \left[\epsilon_k^a
\epsilon_k^b - (V^2 - \Delta_{ab}^2)\right]^2 +\Delta^2
\epsilon_k^{a2} = 0
\end{align}
For the frequency of these excitations to vanish it is required
that $\left[\epsilon_k^a \epsilon_k^b - (V^2 -
\Delta_{ab}^2)\right]^2 +\Delta^2 \epsilon_k^{a2} = 0$. This can
occur by tuning the hybridization parameter, such that,
$V=\Delta_{ab}$ in which case gapless excitations appear at
$k=k_F^a$ where $\epsilon_{k=k_F^a}^a=0$. Without this fine tuning
there are no gapless modes. If, for symmetry reasons, we neglect
the term linear in $\omega$, we obtain the energy of the
excitations in the form $\omega _{12}(k)=\sqrt{\bar{A}_{k}\pm
\sqrt{\bar{B}_{k}}}$ with,
\begin{equation}
\bar{A}_k= A_k + \frac{\Delta^2}{2}
\end{equation}
and
\begin{equation}
\bar{B}_k= B_k +
\frac{\Delta^4}{4}-\frac{\Delta^2}{2}(\epsilon_k^{a2} -
\epsilon_k^{b2})+\Delta^2(V^2+\Delta_{ab}^2)
\end{equation}
where $A_k$ and $B_k$ are given by Eqs. \ref{ak} and \ref{bk}
respectively. In the appropriate limits these equations reduce to
the cases we studied before. Notice that in this case there are
two order parameters in the problem, $\Delta$ and $\Delta_{ab}$,
both defined before. The dispersion relations are shown in Fig.
\ref{fig6}. Excluding the fine tuned case $V=\Delta_{ab}$, any
attractive interaction among the $b$-quasi-particles removes the
gapless modes in the dispersion relations independently of
$\Delta_{ab}$ or Fermi-surface mismatch.
The order parameters are determined by two coupled equations which
for finite temperature are given by,
\begin{align}
\label{gg} & \frac{1}{g
\rho}\!=\frac{-1}{2}\!\int_{-\omega_0}^{\omega_0}\!\frac{d
\epsilon}{\sqrt{B(\epsilon)}}[\left(\frac{\omega_1^2(\epsilon)-
\gamma^2(\epsilon)}{2\omega_1(\epsilon)}\right) \tanh \frac{\beta
\omega_1(\epsilon)}{2} \nonumber \\
&-\left(\frac{\omega_2^2(\epsilon)-
\gamma^2(\epsilon)}{2\omega_2(\epsilon)}\right)\tanh \frac{\beta
\omega_2(\epsilon)}{2}]
\end{align}
and
\begin{align}
\label{gbg} & \frac{1}{g_b
\rho_b}\!=\!\frac{1}{2}\!\int_{-\omega_0}^{\omega_0}\!\!\frac{d
\epsilon}{\sqrt{B(\epsilon)}}[\left(\!\frac{\alpha^2\omega_1^2(\epsilon)\!-\!
(\!\epsilon\!+\!b\!-\!\alpha\!)^2}{2\alpha^2\omega_1(\epsilon)}\right)\!\!\tanh
\frac{\beta
\omega_1(\epsilon)}{2} \nonumber \\
&-\left(\!\frac{\alpha^2\omega_2^2(\epsilon)\!-\!
(\!\epsilon\!+\!b\!-\!\alpha\!)^2}{2\alpha^2\omega_2(\epsilon)}\right)\!\!\tanh
\frac{\beta \omega_2(\epsilon)}{2}]
\end{align}
where
\begin{align}
& \gamma^2=(\frac{\epsilon +(\alpha \epsilon
-b)}{2})^2+(\Delta_{ab}^2-V^2)+\frac{\Delta V}{4}(\Delta V
+ \nonumber \\
& 4(\frac{\epsilon +(\alpha \epsilon -b)}{2})) -(\frac{\epsilon
-(\alpha \epsilon -b)}{2}-\frac{\Delta V}{2})^2
\end{align}
The right hand sides of Eqs. \ref{gg} and \ref{gbg} define the gap
functions $\bar{f}(\Delta, \Delta_{ab})$ and $\bar{f}_b(\Delta,
\Delta_{ab})$, respectively. Adding these equations we get,
$(1/\rho g)+(1/\rho_b g_b)=\bar{G}(\Delta,
\Delta_{ab})=\bar{f}(\Delta, \Delta_{ab})+\bar{f}_b(\Delta,
\Delta_{ab})$. This function is plotted in Fig. \ref{fig7}. For
$\Delta_{ab}\sim V$ and small values of $\Delta$ there is a region
of first order transitions and this remains valid even as $V
\rightarrow 0$. The existence of an intra-band interaction and two
order parameters makes this case qualitatively different from the
pure inter-band interaction even in the limit $V \rightarrow 0$
\cite{kondo}.
\section{Conclusions}

We have investigated superconductivity in two-band systems with
mismatched Fermi surfaces in the presence of hybridization using a
mean-field approximation. For inter-band interactions we found a
phase with gapless excitations on two two-dimensional {\em Fermi
surfaces}. This replaces the BP phase in the case the
quasi-particles can transmute into one another. This phase
corresponds to a metastable minimum of the free energy for a
constant q-independent interaction. Differently from the BP phase
pairing occurs between the Fermi surfaces and this results in a
net gain of energy due to the condensation of these
quasi-particles. In the intra-band case we have shown the
existence of a QCP at which superconductivity is destroyed as
hybridization (pressure) increases beyond a critical value. The
phase diagram and quantum phase transitions can be explored either
by changing the strength of the attractive interactions or the
hybridization. Hybridization among other things varies the
mismatch of the Fermi surfaces. Since in real systems it can be
controlled by external pressure it is a useful parameter to
investigate the effects of Fermi surface mismatch in multi-band
superconductors. Our mean-field approach is more appropriate to
treat weak coupling systems with $g, g_b \sim 1$, although even in
this case it can miss effects due to fluctuations, as an
additional p-wave instability \cite{second}. In the metallic
problem, the quasi-particles have spins as extra degrees of
freedom and in principle there is the possibility of an additional
s-wave pairing between quasi-particles at the gapless Fermi
surfaces. This is taken into account in the mean-field approach
even if the interaction between these quasi-particles is not
included in the Hamiltonian. This manifests through the appearance
of anomalous Greens functions involving these quasi-particles.
However, only in the case $g$ and $g_b$ are finite we find two
order parameters, with none being identically zero.

In heavy fermion materials \cite{livroM, bianchi} hybridization
plays an important role and they could display the effects and
phase transitions discussed above. As hybridization (pressure)
increases giving rise to Fermi surface mismatch, we expect a QCP
associated with vanishing superconductivity for predominant
intra-band interactions.  If inter-band coupling is stronger an
FFLO or other exotic superconducting phases are expected with
increasing hybridization. The origin of the attractive
interaction, whether due to phonons or spin-fluctuations does not
affect the present results, although the use of a mean-field
approximation appears questionable to treat these strongly
correlated materials. However, as pointed out in Ref. \cite{liu},
for fixed $k_F^{a,b}$ and inter-band interactions, the critical
coupling $g^c_{1,2} \rightarrow 0$, as the mass ratio $\alpha
\rightarrow 0$. Since this holds in the presence of hybridization,
HF materials which are characterized by small mass ratios $\alpha$
fall in the weak coupling regime for which the present mean-field
is appropriate.

Multi-band superconductors as $MgB_2$ are also candidates to
investigate the effects discussed here \cite{mgb2}. Pressure
decreases the temperature of the superconducting transition
although in actual experiments in these systems it is not enough
to drive them to a QCP. Evidence of topological electronic
transitions has been found in these experiments. These transitions
involve changes in Fermi surfaces and bear some resemblance
\cite{liu3} with those we studied here. We hope the results
presented in this paper will stimulate further experimental work
in multi-band superconductors.

\acknowledgments The authors thank H. Caldas and A. Troper for
comments and discussions. They also thank the Brazilian Agencies,
FAPERJ, FAPEAM and CNPq for financial support.


\begin{thebibliography}{99}
\bibitem{nota} Throught this work $V$ is sufficiently small, such that, both
quasi-particles states are occupied ($V\lesssim 0.95$).
\end{thebibliography}
\end{document}